\documentclass{aa}

\usepackage[dvips]{graphicx}
\usepackage{natbib}
\usepackage{amsfonts}

\newcommand{\ARAA}{ARA\&A}
\newcommand{\AaA}{A\&A}
\newcommand{\AaAS}{A\&As}
\newcommand{\ApJ}{ApJ}
\newcommand{\MNRAS}{MNRAS}
\newcommand{\Natur}{Nature}

\begin{document}

\def\simgt{\lower.5ex\hbox{\gtsima}}

\title{Separating the kinetic SZ effect from primary CMB fluctuations}

\author {O. Forni \and N. Aghanim }

\offprints{Olivier.Forni@ias.u-psud.fr}

\institute{IAS-CNRS, Universit\'e Paris Sud, B\^atiment 121, F-91405 Orsay Cedex}

\date{Received date / accepted date} 

\abstract{
In the present work, we propose a new method aiming at extracting the
kinetic Sunyaev-Zel'dovich (KSZ) temperature fluctuations embedded in
the primary anisotropies of the cosmic microwave background (CMB). 
We base our study on simulated maps without noise and we consider
very simple and minimal assumptions. Our method essentially
takes benefit from the spatial correlation between KSZ and the Compton
parameter distribution associated with the thermal Sunyaev-Zel'dovich
(TSZ) effect of the galaxy clusters, the later can be obtained by
means of multi-frequency based component separation techniques.  We
reconstruct the KSZ signal by interpolating the CMB fluctuations
without making any hypothesis besides the CMB fluctuations are
Gaussian distributed.  We present two ways of estimating the KSZ
fluctuations, after the interpolation step. In the first way we use a
blind technique based on canonical Principal Component Analysis, while
the second uses a minimisation criterion based on the fact that KSZ
dominates a small angular scales and that it follows a non-Gaussian
distribution. We show  using the correlation between the
input and reconstructed KSZ map that the latter can be reconstructed
in a very satisfactory manner  (average correlation coefficient
between 0.62 and 0.90), furthermore both the retrieved KSZ power
spectrum and temperature fluctuation distribution are in quite good
agreement with the original signal. The ratio between the input
and reconstructed power spectrum is indeed very close to one up to a
multipole $\ell\sim 200$ in the best case. The method presented here 
 can be considered as a promising starting point to identify in CMB 
observations the temperature fluctuation associated with the KSZ effect.

\keywords{Cosmology: Cosmic microwave background - Methods: Data Analysis }}

\maketitle

\markboth{Separating the kinetic SZ effect from primary CMB fluctuations}{}

\section{Introduction}
The Cosmic Microwave Background (CMB) temperature anisotropies contain
the contribution of both the primary cosmological signal, directly
related to the initial density fluctuations, and the foregrounds
amongst which are the secondary anisotropies generated after
matter-radiation decoupling.  They arise from the interaction of the
CMB photons with the matter and can be of a gravitational type
(e.g. Rees-Sciama effect (Rees \& Sciama 1968)), or of a scattering
type when the matter is ionised (e.g. Sunyaev-Zel'dovich (SZ) effect
(Sunyaev \& Zel'dovich 1972) or Ostriker-Vishniac effect (Ostriker \&
Vishniac 1986; Vishniac 1987)). Among all these secondary
anisotropies, the dominant effect is the SZ effect.  It represents the
inverse Compton scattering of the CMB photons by the free electrons of
the ionised and hot intra-cluster gas. It results in the so-called
thermal SZ (TSZ) effect whose amplitude is characterised by the
Compton parameter $y$ (the integral of the pressure along the line of
sight). The TSZ amplitude thus depends only on the cluster electron
temperature and density distributions.  The inverse Compton effect
moves the CMB photons from the lower to the higher frequencies of the
spectrum. This results in a peculiar spectral signature with a
decrement at long wavelengths and an increment at short wavelengths.
When the galaxy cluster moves with respect to the CMB rest frame, {with a 
peculiar radial velocity $v_r$}, the
Doppler shift induces an additional effect often called the kinetic SZ
(KSZ) effect, which generates temperature anisotropies with the same
spectral signature, at least in the non-relativistic approximation,
 as the primary CMB fluctuations.

The interest of the TSZ effect for cosmology has been recognised very
early (see reviews by \cite{rephaeli95}, \cite{birkinshaw99} and \cite{carlstrom02}).
It is a powerful tool to detect high redshift galaxy clusters since it is
redshift independent. In combination with X-ray observations it can be
used to determine the Hubble constant and probe the intra-cluster gas
distribution. Moreover, the KSZ effect may be the one of the best ways
of measuring the cluster peculiar velocities by combining thermal and
kinetic effects \cite[]{suniaev80}.  The advantages of this method
are: (i) it yields directly the peculiar velocities, bypassing the
need to measure inaccurate distance indicators
\cite[]{faber76,tully77}; (ii) the method has a physical 
explanation and (iii) it is independent of distance.  The KSZ can be
distinguished from the TSZ effect due to the different frequency
dependence of their intensities.  The KSZ intensity reaches its
maximum at a frequency of $\sim 218 $ GHz, just where the TSZ
intensity is zero .  Hence, this is the optimal frequency to the
detect the KSZ signal. 
 It has also been shown \cite[]{hobson98,bouchet99,baccigalupi00,delab02,kuo02,maisinger03} that the TSZ signal can be extracted from the other astrophysical 
contribution by component separation techniques (Wiener filtering, Maximum Entropy, 
Independent Component Analysis, ...).  Despite the scientific interest of the KSZ
effect as a probe of large scale matter distribution and structure
formation theories, very few measurements of the peculiar velocities
were achieved \cite[]{holzapfel97,lamarre98,benson03}. 
As a consequence, very few methods have been
proposed so far to address the specific underlying question of {\it
separating the secondary KSZ fluctuations from the primary
anisotropies}. In an early work, \cite{haehnelt96} used an optimal
filtering (Wiener), with a spatial filter derived from X-ray
observations of galaxy clusters, that minimises the confusion with
CMB. However, this method implied the knowledge of the CMB power
spectrum. \cite{aghanim97} rather used a matched filter optimised on
simulated data and independent of the underlying CMB model.
 Recently \cite{hobson03} presented a Bayesian approach for detecting and 
characterising the signal from discrete objects embedded in a diffuse background.
They showed that this approach is around twice as sensitive as the linear optimal filter
approach proposed by \cite{haehnelt96}.

In the present study, we propose a new method optimised to extract
from the primary anisotropies, the temperature fluctuations,
associated with the KSZ effect. The method is based on the fact that
we have two sets of maps (provided, in a realistic case, by component
separation techniques), the first set contains both CMB and KSZ
temperature fluctuations and the second set consists of Compton
parameter maps associated with the TSZ effect which is used as a spatial 
template.  In our study, we
do not use real (observed) maps but we rather use two sets of
simulated maps. We were able to retrieve, in the best possible way,
the amplitude and the distribution of the temperature fluctuations
associated with KSZ together with the associated power spectrum.

\section{Methodology}

In a ``real-life'' case, it is worth noting that the application
of the method we propose here is based on the fact that a first-step
component separation is performed on the CMB data leaving us with a
TSZ effect map and a temperature fluctuation map containing primary
and KSZ anisotropies.  In the present study, we focus on the
description of the method and the way it is intrinsically limited by
the pure cosmological signals primary CMB + SZ effect (without adding
any instrumental effects). It is beyond the scope of this first study
to address the instrumental effects (this will be the subject of a
future work), therefore and as mentioned above, we use simulated 
cosmological data. Namely, we simulate 15 ($512\times 512$ pixels)
primary CMB, TSZ and KSZ maps with a pixel size of 1.5 arc-minutes. A
precise description of the SZ simulations is given in Aghanim et
al. (2001).  The KSZ effect induces temperature fluctuations that
can be written as $\delta_T^{\mathrm{KSZ}}=(\Delta T/T)_{\mathrm
{KSZ}}=-\frac{v_r}{c}\tau$, with $c$ and $\tau$ the velocity of light
and the cluster Thomson optical depth. The primary CMB and the KSZ
anisotropies having the same spectral shape in the 
non-relativistic approximation, we construct maps of radiation
temperature fluctuations, $\delta_T$, by adding the two signals 
$\delta_T=(\Delta T/T)_{\mathrm{KSZ}}+(\Delta T/T)_{\mathrm {CMB}}$.
We are thus left with two data sets { of pure cosmological
signals}, one consisting of the temperature fluctuation maps (CMB +
KSZ) and the other consisting of the Compton parameter maps, $y$, for
the TSZ effect.  For this study, we adopt a low matter density flat
model defined by: $\Omega_{\mathrm m}$ = 0.3, $\Omega_{\Lambda}$ = 0.7
and $h$ = $H_0$/100 km/s/Mpc = 0.65.

Provided the two types of maps, $y$ maps for the TSZ effect 
(mean $\sigma$ = 1.17 10$^{-5}$) and
$\delta_T$ maps for CMB + KSZ, our goal is to obtain the best possible
estimate of the KSZ. To achieve this goal, we benefit from the fact
that the TSZ and the KSZ features are spatially correlated  as
already noted by \cite{diaferio00} and \cite{sorel02}. The later
showed that the absolute values of the covariance coefficients between
TSZ and KSZ maps are significantly high even though the correlation
coefficient between the maps does not exceed 0.1 in absolute value.
This low value is due to the fact the signs and amplitudes of the KSZ
anisotropies in a map depend on the distribution of the radial
peculiar velocities which is a random variable with zero mean.  The
spatial correlation between TSZ and KSZ simply means that both
effects are due to galaxy clusters. Therefore, where the TSZ signal is
present so are the KSZ fluctuations regardless of their signs or
amplitudes. Conversely, where the TSZ fluctuations are absent, so are
the KSZ fluctuations and the signal at that position in the $\delta_T$
map is therefore associated with the CMB anisotropies only.

Our  technique to separate the KSZ fluctuations from the primary
CMB anisotropies is based on this simple statement. It allows us to
build up a two-step strategy in which: (i) we first derive the best
estimate of the CMB map, and (ii) consequently deduce the best estimate
of the KSZ map. 

\subsection{Estimating the primary CMB anisotropies}

In this section, we address the first step of our separation
method, namely we derive an estimate (the best possible) of the
primary fluctuation map. For this we use the two observables: The TSZ
map and the $\delta_T$ map which contains both the primary and the KSZ
fluctuations.  Given the above mentioned statement on the
spatial correlation between TSZ and KSZ signals, the basic idea in
order to estimate the primary CMB anisotropies, is to use the TSZ map
as a mask to select in the $\delta_T$ map, the {\it pixels where the
TSZ fluctuations are not present, i.e. where only primary
anisotropies are present}. The rest of the pixels in the map are
{\it missing or masked pixels}. We then interpolate the $\delta_T$ signal on
these {\it missing pixels} with the constraint that the pixels
where TSZ is absent i.e. with signal associated with primary CMB only, keep
their values after the interpolation is achieved.  We therefore end up
with an estimated primary CMB map where the $\delta_T$ signal in the
masked (missing) pixels is obtained from the interpolation.
Formally, the KSZ map can then be estimated simply by computing the
difference between the original unmasked $\delta_T$ map and the
primary CMB map estimated with the interpolation.

\subsubsection{Interpolation}\label{sec:inter}

We already note that the ``recovery'' of the KSZ map heavily relies on 
the performances of the interpolation method.
We first use the method described in Unser (1995). Consider the problem of
 the minimisation of a general criterion written as: 

\begin{eqnarray}
\label{eq:minsp}
  E(u) = \sum_{(k,l) \in \mathbb{Z}^2} w(k,l) \big[ f(k,l)-u(k,l) \big]^2 +
 \nonumber \\
\lambda \sum _{(k,l) \in \mathbb{Z}^2}\big[d_x*u(k,l) \big]^2 +\big[d_y*u(k,l)
\big]^2 
\end{eqnarray}

\noindent
where $f$ is an input image, $u$ is the desired solution $w \gid 0$ is a map
of space-varying weights, $d_x$ and  $d_y$ are the horizontal and vertical
gradient operators, respectively. The second  space-invariant term in Eq.
\ref{eq:minsp} is a membrane spline regulariser; the amount of smoothness is 
controlled by the parameter $\lambda$.  Taking the partial derivative
of Eq. \ref{eq:minsp} with respect to $u$, we find that $u$ is the
solution of the differential equation :
 
\begin{equation} \label{eq:diffeq}
{f_w}={Wu}+{\lambda L u}={Au} 
\end{equation}

\noindent
where ${W}$ is the diagonal weight matrix, ${{f_w}}= { Wf}$ 
the weighted data vector, ${{ L}}$ 
is the discrete Laplacian operator and ${ A} =  {{ W} +
\lambda { L }}$ a symmetric definite matrix.   
The inversion of Eq. \ref{eq:diffeq} is achieved using a multi-grid
technique (Wesseling, 1992). Typically, we need two V-cycles with two 
iterations in the smoothing Gauss-Seidel part of the algorithm to reach
a residual of the order of $\mathrm{10^{-6}}$.     
In our case, the interpolation of the primary CMB map is achieved 
by setting the weights to zero where
the data are missing, i.e. in the masked pixels, and to one 
elsewhere and by resolving Eq. 
\ref{eq:diffeq}. The value of $\lambda$ then determines the tightness of the
fit at the known data points (unmasked pixels), while the surface 
$u$ is interpolated such that the values of the Laplacian of
$u$ is zero elsewhere. In the present work, we impose a low
value for $\lambda$ so that the recovered values at the known data
points are equal to the original values. This criterion can be relaxed
to take into account corruption of the data by additive white noise
(Unser, 1995). In this case, the optimum regularisation 
parameter $\lambda$ can be defined as:

\begin{center}
\begin{equation}
\lambda ={ {\sigma^2} \over {E(f.Lf)-4\sigma^2}}
\end{equation}
\end{center}

\noindent
where $\sigma^2$ is the variance of the noise and $E(f.Lf)$ denotes an
estimate of the correlation between the noisy image f and its
Laplacian Lf. In the other cases (non white noise), the optimal
regularisation parameter $\lambda$ may be determined from the data
using cross-validation methods \cite[]{wahba77}, or from a given
measurement model of the signal + noise \cite[]{reeves94}.

\begin{figure} %\epsfxsize=\columnwidth
%\hbox{\epsffile{ccor6-12.ps}}
\includegraphics[width=\columnwidth]{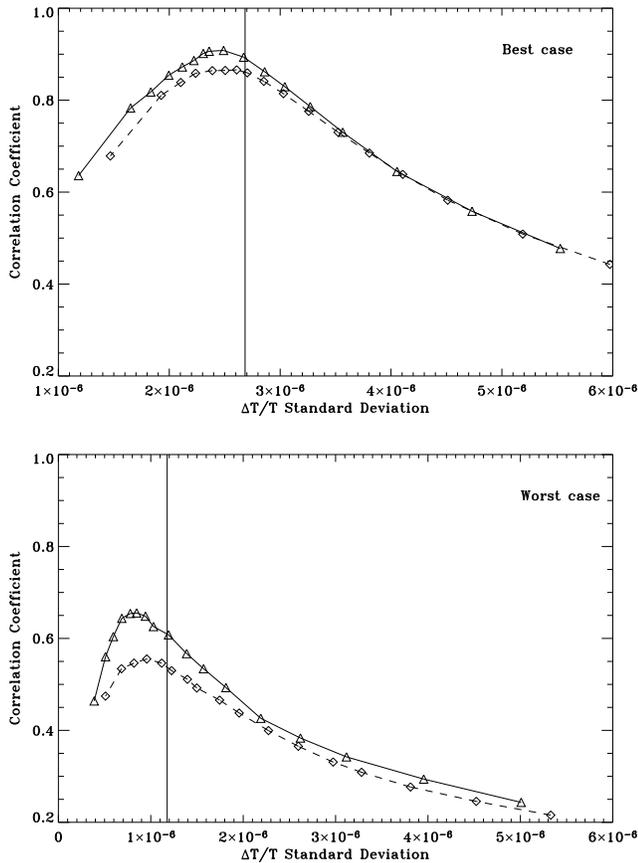}
 \caption{The correlation coefficient between the original KSZ map and
the series of 17 estimated KSZ maps as a function of the
standard deviations of the estimated KSZ maps. Upper panel is for
the best case and lower panel for the worst case. The triangles and
the solid line stand for the interpolation with the biharmonic
operator, the diamonds and the dashed line are for the
interpolation with the Laplacian.  The interpolation with the
biharmonic operator gives better results especially for the KSZ maps
with low standard deviation. The vertical lines mark the standard
deviation of the original KSZ maps ($2.6\: 10^{-6}$ and $1.2\:
10^{-6}$). The standard deviation of the primary CMB is $1.9 \:
10^{-5}$. } \label{fig:cfcor}
 \end{figure} 

\begin{figure}
\includegraphics[width=\columnwidth]{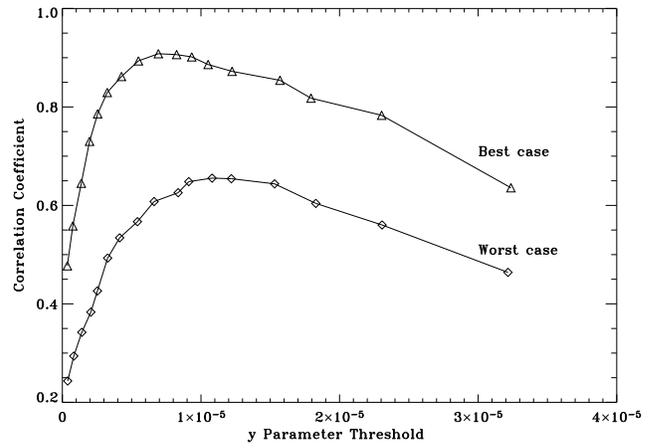}
\caption{For the best case (upper curve and triangles) and worst 
{ case (lower curve and diamonds)}, the correlation coefficient 
between the original and the { 17 estimated} KSZ maps as a function of 
the { associated 17 TSZ} threshold values. The highest threshold value 
is of the order of $y=3.5 \: 10^{-5}$.
The interpolation method uses the biharmonic operator.  } \label{fig:cfcory}
\end{figure}

It is possible to improve the performances of the interpolation, and
hence of the retrieved KSZ map, by setting non-zero values to the {
Laplacian of \it{u}} at the missing data points (which are set to zero
in the original method). The values we set for the { Laplacian of
\it{u}} are such that the first and second derivatives { of the
interpolated signal} are continuous throughout the interval.  These
continuity conditions characterise the cubic B-spline functions which
are known for their simplicity and their performances in terms of
signal reconstruction (Unser et al., 1993; Th\'evenaz et al. 2000).
In practice, these additional conditions imply that the source term
{{$f_w$}} in Eq. \ref{eq:diffeq} is modified to impose non-zero
values at the points where the weights are set to zero (i.e. the
missing data points). An equivalent way to solve Eq. \ref{eq:diffeq}
with the above mentioned conditions, is to replace the Laplacian
operator ${L}$, by the quadratic operator ${L^{2}}$. { These two
interpolation methods are obviously not the unique techniques and
other techniques (based on textures for example) can be used. In the
following we will only test the two operators described above and then
choose among them, the one which gives the most satisfying results.}

\subsubsection{Defining the mask or missing pixels}\label{sec:mask}

We must now define more precisely what we mean when we state {\it
where the TSZ is not present}; or in other terms how do we select the
missing data points? Besides the pixels that actually contain no
galaxy clusters, i.e. no SZ contributions, this statement means that
we fix a threshold value for the TSZ amplitude below which we consider
the TSZ signal is { too small to be detected}. The corresponding
pixels in the $\delta_T$ maps are then considered to be associated
only with the primary CMB signal. On the contrary, above this
threshold the corresponding pixels in the $\delta_T$ map are
considered to be the missing data points, i.e. masked pixels that we
want to interpolate. It is clear that the number and location of the
missing data will depend on the threshold.  The lower it is, the
larger the number of missing data we need to recover.  The choice of
this threshold has also important consequences on the quality of the
interpolation.

When the threshold is high, on the one hand, the number of missing
data is small and the interpolated surface is good. On the other hand,
the selection retains only the clusters with the highest TSZ and
misses the majority of clusters. In this case, we expect to end up
with a low correlation coefficient between the retrieved and the
original KSZ maps.  When the threshold is low, we take into account a
majority of clusters, but the interpolated surfaces are large and the
quality of the interpolation suffers from that. Moreover, the
characteristic scale of the interpolated surfaces becomes, in this
case, of the order of that of the CMB fluctuations, leading to
``confusion effects'' in the interpolation.  From these remarks, we
can infer that: Firstly, there will exist an optimal threshold value
for which the correlation between the retrieved and the original KSZ
maps is maximum. Secondly, the restoration of extended clusters is
likely to be of low quality as already noted by \cite{haehnelt96}.

Obviously there is no a priori way of choosing the TSZ threshold on an
objective basis. Indeed, given the TSZ map is obtained (in
``real-life'') from a component separation process involving the true
signal but also the instrumental and observational effects, one cannot
rely on a ``theoretical'' expected value.  Therefore, rather than
performing only one interpolation of { the primary CMB map} for one
single { TSZ} threshold, we propose to retrieve a
set of interpolated CMB maps corresponding to a set of TSZ threshold
values. The later { can be defined in a simple way
without any theoretical or observational prior} as follows: We compute
the cumulative distribution function of the TSZ values in the given
map and we search for the values corresponding to 5\% to 95\% of the
total number of pixels (with a step of 5\%).  This gives us a set of
19 threshold values such that all pixels in the TSZ map that have $y$
parameters above the threshold are identified as the missing data
points in the simulated $\delta_T$ map, i.e. the mask. { In the present
study, we are using simulated TSZ maps, i.e. without noise. These maps
exhibit} a background of zero values which proportion represents in
our case at least 10\% of the total pixel number. This characteristics
implies that the first { two TSZ threshold values associated with 5
and 10\% of the total pixel number} are irrelevant.  { In the following, we
thus use only the 17 highest TSZ thresholds}. 

\subsubsection{Results}

{ For each of the 15 simulated maps, we obtain 17 TSZ threshold values, 
and thus 17 masked
$\delta_T$ maps. We apply the interpolation techniques
(Sect. \ref{sec:inter}) to recover the primary CMB signal in the
masked regions. For each of the 15 simulated maps, we thus end up 
with 17 estimated primary CMB maps
corresponding to the 17 TSZ threshold values. The associated KSZ maps
are evaluated simply by subtracting the interpolated primary CMB maps
from the total $\delta_T$ map.

In order to evaluate how well the two interpolation methods presented in
Sec. \ref{sec:inter} do recover the KSZ signal, we compute for
each of the 17 KSZ estimated maps the correlation coefficient between
the original input KSZ map and the estimated KSZ maps.  The results
are shown in Fig. \ref{fig:cfcor}. The data points representing the
correlation coefficients are plotted as a function of the standard
deviation of the estimated KSZ map for each of the 17 threshold
values.  The diamonds and the dashed line represent the case where the
interpolation is such that the Laplacian values are set to zero, and
the triangles and the solid line are for the case in which the
Laplacian values are non-zero.  The upper panel in
Fig. \ref{fig:cfcor} shows our best recovery case in terms of
correlation coefficient. The lower panel is for our worst case.  The
correlation coefficients between the original input KSZ map and the 17
estimated KSZ maps are also displayed as a function of the 17 TSZ
threshold values in Fig. \ref{fig:cfcory}. It is worth noting that the
high TSZ thresholds (abscissae in Fig. \ref{fig:cfcory})
correspond to low KSZ standard deviations (abscissae in
Fig. \ref{fig:cfcor}).}

First, we note { from Fig. \ref{fig:cfcor}} that for any standard
deviation of the estimated KSZ map, the correlation coefficient
between the original and the estimated KSZ maps is higher when {
the Laplacian values are non-zero than when they are set to zero.
Actually there can be a significant improvement in the KSZ
reconstruction if an optimised interpolation method is used.} This is
especially true for the maps with low standard deviations. { The
improvement brought by the biharmonic operator is of the order of
20\% in our worst case (Fig. \ref{fig:cfcor}, lower panel)}. We will
therefore use, in the following, the most powerful interpolation 
method { that is the one with the $L^2$ operator}.

Second as expected, the correlation coefficient increases when the TSZ
threshold decreases as shown in Fig. \ref{fig:cfcory} (i.e. when the
standard deviation of the estimated KSZ map increases in
Fig. \ref{fig:cfcor}). { The correlation coefficient reaches} a
maximum value and then it decreases for the lowest TSZ thresholds
(i.e. the highest { KSZ} standard deviations). Moreover, we note
that { among the set of 17 KSZ estimated maps the one} with the
highest correlation coefficient is { shifted towards lower values
proportionally to the standard deviation of the original KSZ map.} We
will use this behaviour later on in the minimisation procedure.

\subsection{Reconstructing the KSZ map}

In the previous step, we { have interpolated the $\delta_T$ signal
to estimate the primary CMB map and then extract the KSZ signal as a
function of a set of TSZ thresholds}. We have tested the performances
of two interpolation methods by comparing { through a correlation
coefficient each of the 17 estimated KSZ maps, corresponding to the 17
TSZ threshold values, to the original input KSZ map} which, of course, we
do not have in ``real life''. { The set of 17 estimated KSZ maps
were obtained by subtracting the interpolated primary CMB maps from
the total $\delta_T$ map of the temperature fluctuations.}

In this step, using the set of { 17 KSZ} estimated maps
associated with the set of { 17 TSZ} thresholds, we search for a
method that gives us either the reconstructed KSZ map which is the
closest to the original KSZ signal or even better, the combination of
the set of KSZ maps giving the best estimate of the original KSZ map.
In the following, we have explored two ways to achieve this goal
(restricted to linear combinations only). The first one is to
decorrelate the set of images by a canonical Principal Component
Analysis (PCA), the second way is to minimise a criterion, which in
our case is related to the non-Gaussian character of the KSZ signal.

\subsubsection{Decorrelation with Principal Component Analysis}

It is obvious { from our definition of the masked pixels
(Sec. \ref{sec:mask})} that all the interpolated maps (defined by the
set of { 17 TSZ} thresholds) are highly correlated. The first and
natural approach to decorrelate them { is thus to use} a PCA
method.  As noted
in Sec. \ref{sec:mask}, decreasing the TSZ threshold increases the
characteristic scale of the structures we have to interpolate. This
means that the confusion between the extended clusters and the {
primary} CMB anisotropies increases.  As a consequence, decreasing the
{ TSZ} threshold increases the proportion of the $\delta_T$ signal
due to the primary CMB in the { estimated KSZ} maps. 
In this context, the purpose of performing a PCA is to
decorrelate the signal due to the primary CMB from that due to the
galaxy clusters. We expect to find the high frequency part of the KSZ
effect in one { principal} component, and in a second {
principal} component, the low frequency part of the KSZ signal
(essentially the extended clusters) together with the contribution
from the primary CMB fluctuations.

{ For each of the 15 simulated maps, we apply} the PCA to the set
of { 17 estimated} KSZ maps. We find that the first and second {
principal} components represent respectively $\sim$ 70\% and $\sim$
15\% of the total input signal. { In our approach, it is the first
principal component which stands for the KSZ reconstructed signal. It
is thus interesting to evaluate how well the PCA performs the
reconstruction. To do so, we compute for each of the 15 simulated
maps} the correlation coefficient between the first { principal}
component and the original input KSZ map. { The correlation
coefficient averaged over the 15 maps reaches 0.73 which is
satisfactory. However, the standard deviation of the first principal
component is on average smaller by almost 50\% than the standard
deviation of the original KSZ signal. The PCA method clearly
underestimates the reconstructed KSZ signal which is an obvious
weakness of the method. We will thus investigate in the following
minimisation methods.} 

\subsubsection{Statistical minimisation}  

By minimising on the known KSZ signal { (from our dataset of 15
input maps), we can first search} for a linear combination of the
set { of 17 estimated KSZ maps} that is the closest to each original
KSZ in the sense of least squares.  This has been done using a
standard singular value decomposition \cite[]{press92}.{ We compute
again the correlation coefficient between the original KSZ map and the
reconstructed map obtained from the minimisation to estimate the power
of the method}. We find an average correlation coefficient { (over
the 15 simulated input maps)} of 0.8, only slightly higher than {
the PCA result of 0.73.} However, the standard deviations of the {
reconstructed} maps are again significantly lower than that of the original
KSZ maps by almost 25 \% { on average (better than in the PCA
case)}.  Furthermore, the results { of the least square
minimisation} depend strongly on the set of estimated maps that are
used which is clearly undesirable.

In order to avoid this problem and to obtain as more map-independent
results as possible, we must identify a trustful criterion to minimise
on. { The latter} should ideally give at the same time a result
that is the closest possible to the { largest} correlation
coefficient of 0.80 { on average (obtained with the least square
minimisation), and reconstructed KSZ maps with the closest possible
standard deviations to those of the original KSZ signal. Moreover,} a good
minimisation criterion  would be a criterion that characterises the KSZ signal
{\it only}, excluding the primary CMB signatures.  We have identified
two properties of the KSZ { fluctuations} that fulfill this definition:

\begin{itemize}
\item The KSZ signal dominates the primary CMB at high wave numbers
(small angular scales).
\item The KSZ effect is a highly non-Gaussian process contrary to the 
primary CMB which is a Gaussian process. 
  \end{itemize} 

\begin{table*}
\begin{center} 
\begin{tabular}{|c||c|c|c|} 
\hline
  & Standard deviation & Skewness &  { Excess kurtosis} \\ \hline \hline
KSZ+CMB &  6.45 $10^{-7}$ &  0.10 & 8.71  \\
KSZ &  6.45 $10^{-7}$ &  0.10 & 8.72  \\ \hline
KSZ+CMB &  2.05 $10^{-7}$ &  0.22 & 8.97  \\
KSZ & 2.09 $10^{-7}$ &  0.23 & 9.15  \\ \hline
CMB & 1.60 $10^{-8}$ &  -0.02 & 0.45  \\ \hline 
\end{tabular} 
\end{center}
\caption{The statistical properties of the first scale { (3 
arc-minutes)} diagonal
wavelet coefficients distribution for the $\delta_T$ map (KSZ + CMB), the
KSZ map and the primary CMB alone.
The two cases stand for our best case (first pair) and the worst
case (second pair). We note that the three moments are almost identical and
characterise well the KSZ fluctuations; they are very different from 
the CMB fluctuations properties.  } \label{tab:stat} 
\end{table*}

\begin{figure} 
\includegraphics[width=\columnwidth]{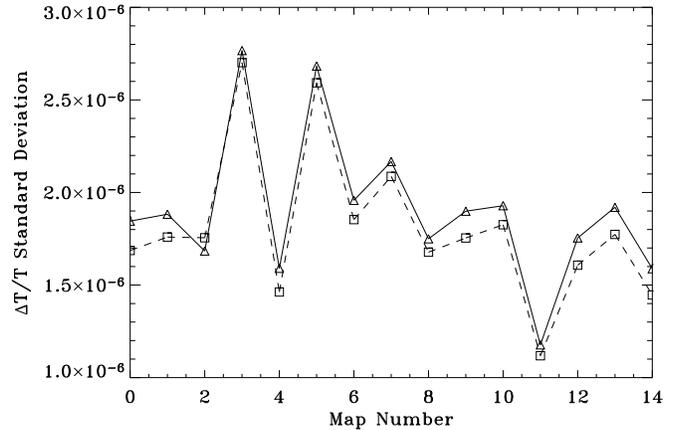}
\caption{Standard deviations of our set of 15 KSZ { original} simulated 
maps { (triangles) as compared to the standard deviations of the 15
reconstructed KSZ maps (squares). The reconstruction is based on the 
minimisation of the statistical criterion (see text).} } \label{fig:dev}
\end{figure} 

The analyses of the available CMB data
\cite[]{cayon03,komatsu03,santos03} all seem to agree on the fact that
primary CMB anisotropies are Gaussian distributed as expected from the
simplest inflationary models. By contrast, the SZ effect is definitely
characterised by its non-Gaussian signatures.  Using wavelet analysis,
we have demonstrated (Aghanim \& Forni 1999; Forni \& Aghanim 1999),
that the excess kurtosis of the wavelet coefficients allows us to
discriminate between a Gaussian primary CMB signal and a non-Gaussian
process like the SZ effect. { We note though that a recent
re-analysis of the WMAP data (Vielva et al. 2003) suggests that the
observed CMB anisotropies exhibit non-Gaussian signatures. The
analysis indicates that the deviations from Gaussianity concern large
scales (a few degrees). Such a behaviour if confirmed would therefore
not affect our minimisation criteria since the latter is based on the
characteristics of the signal at small angular scales (a few to a few
tens of arc-minutes).} In particular, the statistical properties of
the wavelet coefficients at the lowest decomposition scale { (3
arc-minutes)} reflect the properties of the SZ effect only. This is
due to our choice of the wavelet basis and of the decomposition scheme
which focus on the scale where the SZ signal dominates over the
primary CMB. Within this choice, the wavelet analysis provides us with
the wavelet coefficients associated with diagonal, vertical and
horizontal details in the analysed map. Finally, we have shown in
Aghanim \& Forni (1999) that in the case of the SZ effect the diagonal
details are by far the most sensitive to the non-Gaussian signatures
(recently confirmed and explained by Starck et al. (2003)).

In Table \ref{tab:stat}, we compare{, using the 9/7 bi-orthogonal filter bank
of \cite[]{cohen90} and for our worst and best cases}
the statistical properties { (standard deviation, skewness and
excess kurtosis)} of the diagonal details of the KSZ maps and CMB+KSZ
maps at the first decomposition scale { (3 arc-minutes)}. { We
also give the values of the three quantities for the primary CMB maps.}
We immediately note that both sets of wavelet coefficients { for
KSZ and KSZ + CMB} share the same statistical properties { and are
quite different from those of the primary CMB alone. This confirms
that not only} the KSZ signal dominates over the primary CMB (same
{ standard deviation}, i.e. same power), { but also} that the
non-Gaussian signatures { in the KSZ + CMB maps} are associated
with the KSZ effect { (same skewness and excess kurtosis). The
above mentioned} properties characterise, { in a mixture of CMB + KSZ 
fluctuations,} the KSZ effect {\it
only}. Consequently, we can confidently minimise on them. In practice,
we choose the following criterion :

\begin{equation} \label{eq:min}
\zeta=Min \bigl[{{({\cal M}_ 2(w_0)- {\cal M}_2(w))^{2}} \over {{\cal M}^{2}_ 2(w_0)}} + {{({\cal M}_4(w_0)- {\cal M}_4(w))^{2}} \over {{\cal M}^{2}_ 4(w_0)}} \bigr]
\label{eq:mincrit}
\end{equation}

\noindent
 { where $w_0$ is the
distribution of the diagonal wavelet coefficients for the known
$\delta_T$ map (KSZ + CMB)} and $w$ is the distribution of the diagonal wavelet 
coefficients for the desired solution { map}.  ${\cal M}_2$ and ${\cal M}_4$ are
respectively the second and the fourth moments of the wavelet
coefficients. This criterion has { thus} the advantage of taking
into account both the energy (or power) content of the coefficients,
through the second moment, and the non-Gaussian character, through the
fourth moment.  We have chosen the fourth moment to characterise the
non-Gaussian property because it is the moment for which the KSZ
signal is the most sensitive to non-Gaussianity { as shown by the
hydro-dynamical simulations of \cite{dasilva01}}.  Clearly, we {
might} also include the third moment { of the wavelet coefficients
to the criterion. This would be needed in particular if we were
dealing with a ``skewed'' signal such as distorted CMB anisotropies by
the weak lensing of large scale structures.  Taking the fourth moment
in the minimisation criterion allows us in turn to focus on the
reconstruction of KSZ maps excluding the signal that might contribute
at that particular angular scale.} { In our minimisation criterion,
we have added the second and fourth moments quadratically. We have
thus attributed equal weights to the power and to the non-Gaussian
character of the KSZ signal.  It is possible to envisage a different
weighting of one term or the other in Eq. \ref{eq:mincrit}. This would
in principal enhance the non-Gaussian signal, for example, and thus
ease its separation from a Gaussian signal. Such a non-quadratic
mixture would be particularly useful at scales where the KSZ effect is
not the dominant process.  However, there is a priori no trivial way
of setting the weights. This possibility should be investigated in the
future.}

The { solution} map $w$ is obtained by minimisation { of the
criterion $\zeta$} over all the combinations { of wavelet
coefficients $w_0$} allowed by our set of { 17 estimated KSZ maps (there
are 15 simulated input maps)}. However, there are far too many
combinations and we therefore choose to reduce the number of cases. To
do so, we recall the observation made in the previous section that the
correlation coefficients between the { original KSZ map and the 17
estimated KSZ maps} exhibit a maximum value (see
Fig. \ref{fig:cfcor}). { In order to reduce the number of
combinations for the minimisation, we} then adopt the following
iterative strategy: At each step, we first eliminate the {
estimated KSZ} map which contains the highest contribution from the
CMB, i.e. the one with the highest standard deviation. Then we
minimise { $\zeta$} using the remaining set of { estimated}
maps. Finally, we take { as a solution of the minimisation
criterion} the map corresponding to { its lowest value}.  

{ In addition to the previous conditions (power and non-Gaussian
character),} we also make use in the minimisation process, of a nice
property of the wavelet transform, which is it preserves the spatial
information.  In our case, this means that we can identify the
diagonal wavelet coefficients that are spatially associated with the
clusters in the TSZ map. Thus instead of minimising over all the {
wavelet coefficients of the data map ($w_0$ in Eq. \ref{eq:min})}, we
can minimise only over the coefficients corresponding to the
clusters. This has two advantages; the first is to enhance the
non-Gaussian character and the second is to reduce the influence of
other possible non-Gaussian processes that could affect the anisotropy
map $\delta_T$. 

\begin{figure}
\includegraphics[width=\columnwidth]{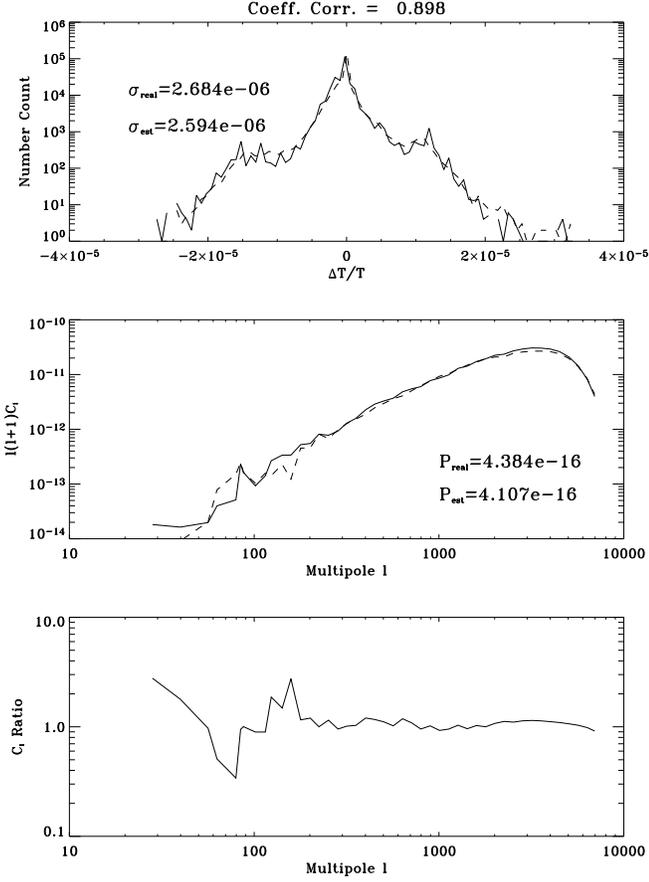}
\caption{Top and middle panels: Histogram and power spectrum of the 
original KSZ map (solid line), and of the reconstructed KSZ map (dashed 
line). { The reconstruction is based on the minimisation of the
statistical criterion (see text).} The bottom panel
exhibits the ratio of the two power spectra. Note the correlation 
coefficient { between the original and reconstructed KSZ maps of $\sim$ 0.9}
and the total power {$P_{\mathrm real}$} and {$P_{\mathrm est}$}}
\label{fig:hist6} 
\end{figure} 

\begin{figure}
\includegraphics[width=\columnwidth]{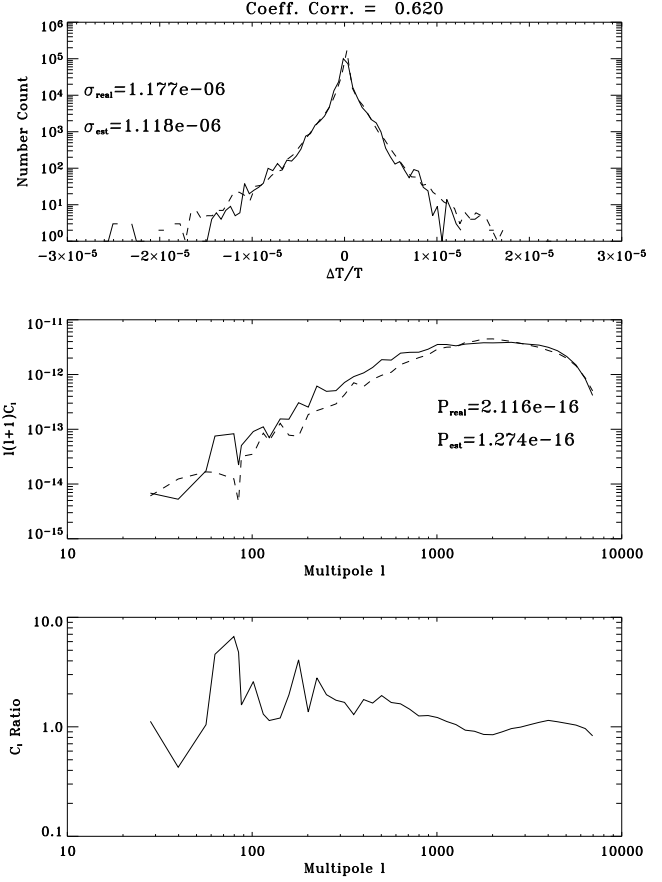}
\caption{Same as figure \ref{fig:hist6}. This is our worst case and 
corresponds to the original KSZ map with the lowest standard deviation. 
{ Note the low correlation coefficient 0.62} }
 \label{fig:hist12} 
\end{figure}

% \begin{table*} %\begin{center} %\begin{tabular}{|c||c||c|c||c|} %\hline
%Method & Corr. Coef.  &  $\sigma(KSZ0 - KSZ)$ & $\sigma_{KSZ0} -\sigma_{KSZ}$ \\
%\hline %SVD  & 0.80 &   1.09 $10^{-6}$ & 3.36 $10^{-7}$   \\
%Mean & 0.71 & 1.38 $10^{-6}$ & 1.21 $10^{-7}$  \\
%PCA (first component) & 0.73 & 1.27 $10^{-6}$ & 4.80 $10^{-7}$  \\
%Max. Coef. & 0.77 & 1.19 $10^{-6}$ & 1.66 $10^{-7}$   \\ %\hline
%\end{tabular} %\end{center}
%\caption{ Comparison of the three methods (SVD, Mean, PCA, Max PCA Coef) 
%using the same interpolated
%maps set with the original KSZ (KSZ0). The first line corresponds to the 
%reference result using a SVD minimisation to retrieve the closest solution.
%The first column represent the mean correlation coefficient. The second column
%displays the mean standard deviation of the difference between the predicted
%KSZ and the original KSZ. The third column stands for the mean difference
%between the standard deviations of the predicted and original KSZ.     %  }
%\label{tab:pca} %\end{table*}

\begin{figure}
\includegraphics[width=\columnwidth]{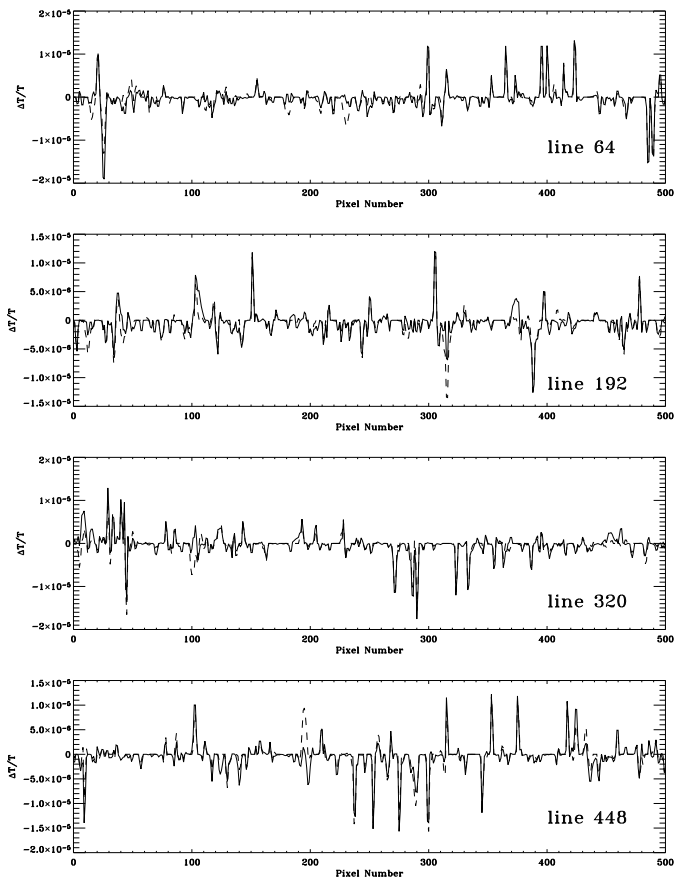}
\caption{{Cuts across the best reconstructed KSZ map (dashed line) and 
its original counterpart (solid line). The cuts have the same 
position in both maps. }} \label{fig:cut6} 
\end{figure}

\begin{figure}
\includegraphics[width=\columnwidth]{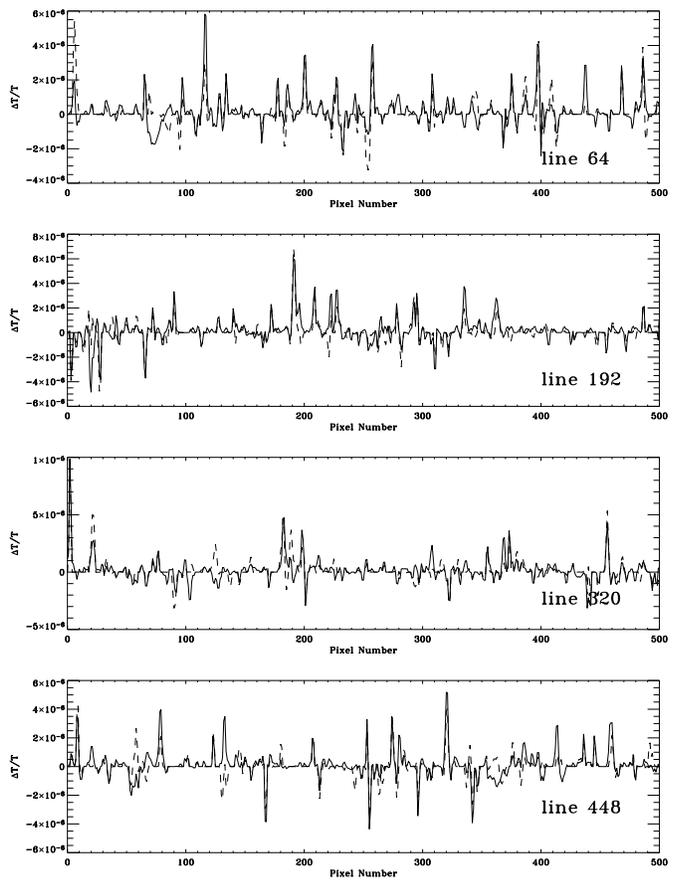}
\caption{{Same as figure \ref{fig:cut6} for the worst reconstructed KSZ 
map and its original counterpart. }} \label{fig:cut12} 
\end{figure}

In Fig. \ref{fig:dev}, we present the standard deviations of the {
15 original simulated KSZ maps (triangles) and of the 15 reconstructed
KSZ maps (squares)} obtained by the minimisation technique described
above. The agreement is pretty good { even for the maps with the
lowest standard deviations. We find the error on the standard
deviation is only of the order of $\sim$5\%. This is much smaller than
what was obtained from the PCA method ($\sim$ 50 \%) or from the least
square minimisation method ($\sim$25\%).  Furthermore, the mean value
(over the 15 original maps) of the correlation coefficient between the
original and the reconstructed KSZ maps} is 0.78. It is slightly
better than the value obtained with the PCA method and quite close to
the value of 0.80 obtained with the least square method.  The quality
of the KSZ map reconstruction can be observed in Figs. \ref{fig:hist6}
and \ref{fig:hist12} which display, for our best and worst cases {
respectively}, the histograms of the temperature fluctuations and the
power spectra of both the original (solid line) and reconstructed
(dashed line) KSZ maps as well as the ratio of these two power
spectra.  It is worth noting that the ratio is close to one over a
large range of multipoles (angular scales) even in the domain where
the primary CMB dominates the KSZ signal by orders of magnitude. {
We also notice the correlation coefficients between the original and
the reconstructed KSZ maps which reaches $\sim0.9$ in our best case
and 0.62 in our worst case. The comparison between the standard
deviations of the original and the reconstructed map $\sigma_{\mathrm
real}$ and $\sigma_{\mathrm est}$ also gives a global indication on
how well the method works.}  Clearly, the method we propose { to
separate between the KSZ signal from the primary CMB anisotropies
despite their identical frequency dependence} allows us to obtain such
results because we were not only able to estimate correctly the
amplitude of the KSZ signal for most clusters but also their {
angular} separation as well as the amplitude of the background
(primary CMB). This is nicely exhibited by the superposition of the
cuts across the reconstructed { (dashed line)} and the original
{ (solid line)} KSZ maps, once again for the best and worst cases
(Figs. \ref{fig:cut6} and \ref{fig:cut12} { respectively}). {
The method partially fails to find broad KSZ features due to their
important level of confusion with the primary CMB fluctuations
(Sec. \ref{sec:mask}).} { Moreover, since the minimisation process
is an overall procedure, it can occasionally happen that relatively
large features (i.e of the order of $10^{-5}$ in absolute $\Delta
T/T$) are poorly recovered}.

\section{Sensitivity test}\label{sec:sens}

{ We have tested our method to separate the KSZ anisotropies
from the primary CMB signal on simulated maps free of any noise.
Moreover, we did not take into account other astrophysical
contributions than the CMB and the SZ effect themselves. In
``real-life'', the data are corrupted by instrumental noise and
astrophysical signals. Additional noise (whatever its origin) will
have a first effect of reducing the ratio between the primary CMB and
the KSZ signals.}  We test the performances of our method { to this
effect} by applying our procedure to one same KSZ map { that is
added to the same primary CMB map. The standard deviation of the KSZ
signal is reduced while the CMB standard deviation is kept} the same
CMB map. { This results in} lowering the KSZ contribution to the
$\delta_T$ map. We { arbitrarily} choose to reduce the standard
deviation { following a} geometrical progression $\sigma_{i}=
\sigma_{0}{\sqrt 2}^i$ with $i=0,6$ and $\sigma_{0}$=2.5
$10^{-7}$. The highest standard deviation is then $\sigma_{max}$=2.0
$10^{-6}$ { which is a typical value for our dataset (see
Fig. \ref{fig:dev})}.

{ We also} test the sensitivity of our method to the wavelet transform 
entering in the minimisation criterion { by comparing} 
the results obtained using two different bi-orthogonal wavelet bases,
the commonly used 9/7 tap filter \cite[]{cohen90} 
and the 6/10 tap filter given by Villasenor et al. (1995). 
\begin{table*}
\begin{center} 
\begin{tabular}{|c||c|c||c|c|} \hline
 & \multicolumn{2}{c||}{9/7 filter} & \multicolumn{2}{c|}{6/10 filter} \\
original $\sigma$ & estimated $\sigma$ &  correlation coefficient & estimated
 $\sigma$ &  correlation coefficient \\ \hline
 2.5 $10^{-7}$ & 2.21 $10^{-7}$ & 0.48 & 2.68 $10^{-7} $ & 0.45 \\
 3.53 $10^{-7}$ & 3.36 $10^{-7}$& 0.54 & 4.02 $10^{-7} $ & 0.52 \\
 5.0 $10^{-7}$ & 5.52 $10^{-7}$& 0.56 & 5.46 $10^{-7} $ & 0.58  \\
 7.07 $10^{-7}$ & 8.02 $10^{-7}$& 0.59 & 6.60 $10^{-7} $& 0.67  \\
 1.0 $10^{-6}$ & 9.74 $10^{-7}$ & 0.68  & 9.16 $10^{-7} $ & 0.71 \\
 1.41 $10^{-6}$ & 1.35 $10^{-6}$ & 0.74 & 1.20 $10^{-6} $ & 0.77 \\
 2.0 $10^{-6}$ & 1.94 $10^{-6}$& 0.78 & 1.77 $10^{-6}$ & 0.81 \\ \hline
\end{tabular} 
\end{center}
\caption{Standard deviations { of the KSZ maps} and correlation 
coefficients { between original and reconstructed KSZ maps} for 
the same KSZ map with standard deviations ranging from 2.5 $10^{-7}$ to 
2.0 $10^{-6}$. Two wavelet bases are tested.} \label{tab:sens}
\end{table*}

 \begin{figure} \includegraphics[width=\columnwidth]{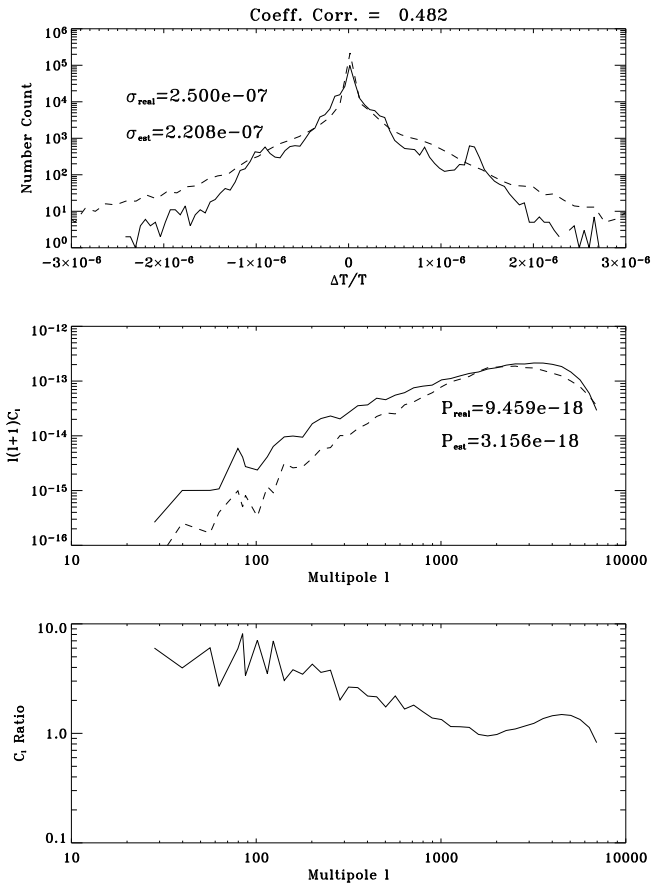}
\caption{Top and middle: Histogram and power spectrum of the original
KSZ map (solid line), and the reconstructed map (dashed line). Bottom
panel exhibits the ratio of the two power spectra. { The standard
deviation of the original map is very low ($\sigma_{0}$=2.5
$10^{-7}$). Note the excess of near zero values in the histogram of
the estimated map (logarithmic scale).  Note also the very low
correlation coefficient 0.48}. This is for the worst case. }
\label{fig:sens1}
\end{figure} 

\begin{figure}
\includegraphics[width=\columnwidth]{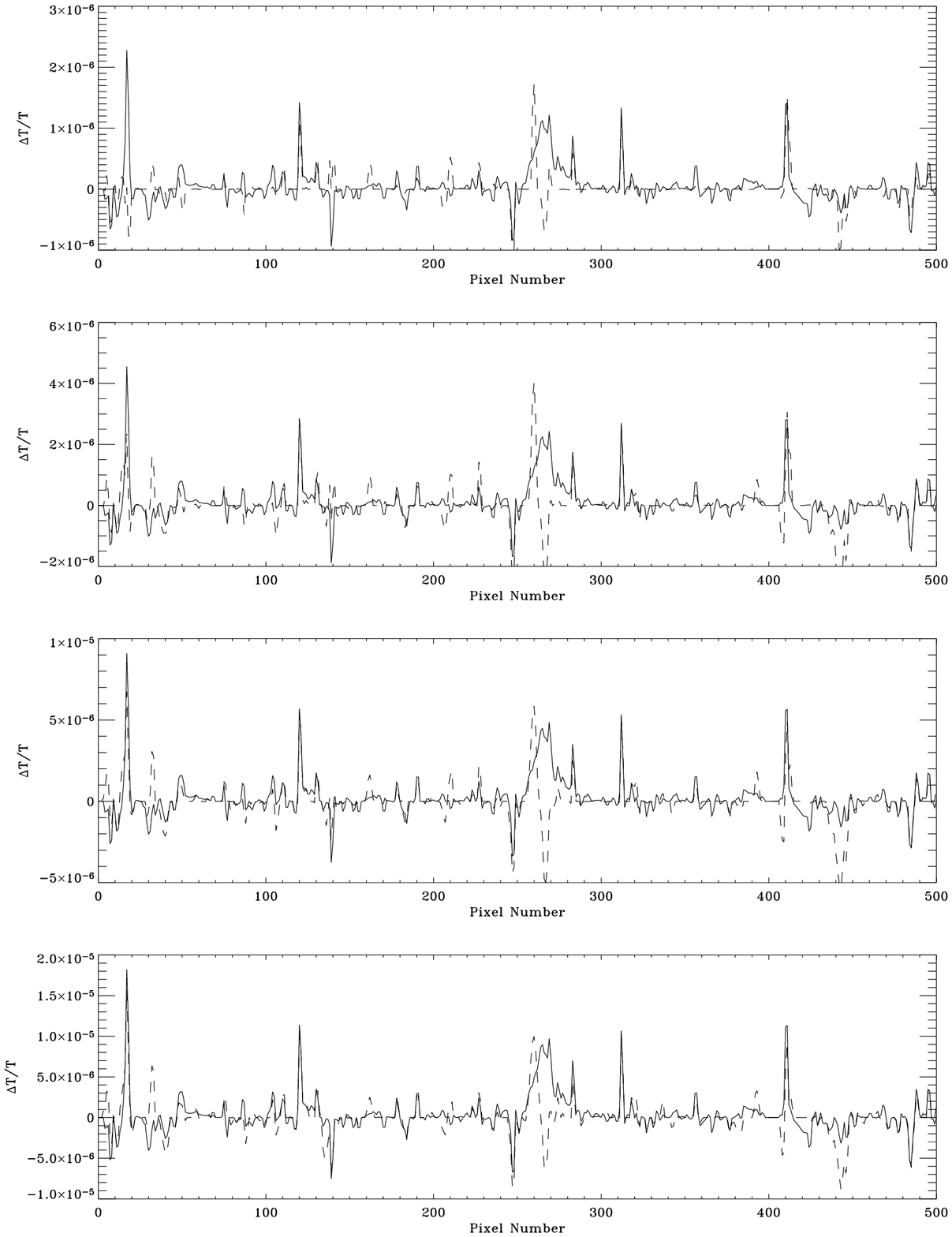}
\caption{Cuts across { original (solid line) and reconstructed
(dashed line) KSZ maps}, at the same position, but with{ increasing
(from top to bottom)} original standard deviation (for $\sigma$: 2.5
$10^{-7}$, 5.0 $10^{-7}$, 1.0 $10^{-6}$ and 2.0 $10^{-6}$). The
important difference { between original and reconstructed signals}
in the middle of the cuts illustrates the fact that large scale
structures (this one is $\simeq 50$ arc-minutes) are poorly resolved
due to confusion with CMB fluctuations. Note that the largest
fluctuation in the left is not reconstructed { for the lowest
$\sigma$} and that the quality of its reconstruction increases with
increasing $\sigma$. }
\label{fig:sens2} 
\end{figure}

The results for this new set of maps are displayed in Table
\ref{tab:sens} in terms of { the standard deviations of the
original and reconstructed KSZ maps,} and of the correlation
coefficient between the original and reconstructed KSZ maps.  We first
notice that the results do not depend much on the wavelet basis. As
expected, the quality of the reconstruction (given in terms of the
correlation coefficient) increases with the standard deviation of the
original KSZ map { from 0.5 to $\sim0.8$. The smallest coefficients
are obtained for very low standard deviations ($<10^{-6}$).} For the
{ KSZ map with the} lowest standard deviation, the histogram of the
reconstructed map { (Fig. \ref{fig:sens1}, upper panel, dashed
line)} shows that the smallest temperature fluctuations are not
resolved, which produces an excess of zero values. More generally,
{ the histogram of the reconstructed map} behaves like a global
envelope to the original histogram { (Fig. \ref{fig:sens1}, upper
panel, solid line)} which does not resolve the details, e.g. the
excess of points around $\Delta_T/T$ = 1.3 $10^{-6}$. { In
addition, we not an overall raise of the wings of the distribution.}
Also the power spectrum computed from the reconstructed KSZ map{
(Fig. \ref{fig:sens1}, middle panel, dashed line)} presents an excess
of power { as compared to the original power spectrum} around
$\ell=2000$ due to the spatial distribution of the unresolved
fluctuations.  The latter also causes the lack of power at higher
multipoles. This behaviour can be also observed in figure
\ref{fig:sens2}, which { shows a cut} across the KSZ maps {
(solid line for the original signal, and dashed line for the
reconstructed signal,)} at the same position but for different
standard deviations. { It can be noticed that the large scale
feature $\sim$ 50 arc-minutes wide at the centre of the cut is poorly
resolved due to confusion with the CMB fluctuations. In that worst case, the
amplitude of the estimated signal remains, however, proportional to the input
signal, but generally the estimation becomes better with increasing standard deviation
of the input signal, as it can be seen for the $\sim$ 10 arc-minutes wide fluctuation
at the left part of the cut.}

\section{Discussion}

{ We present a method for separating the KSZ signal from primary CMB
anisotropies based on two steps: 1) Interpolation and 2) reconstruction}.
Our results clearly depend on the quality of the interpolation {
used to estimate the primary CMB signal and thus the KSZ maps}. In our
case this corresponds to the interpolation of a correlated noise,
namely the CMB.  The results we present in this study seem already
very satisfactory but might certainly be improved. 

The KSZ reconstruction is based on the set of KSZ estimated maps
obtained with a specific choice of TSZ thresholds. We have used here a
rather simple but robust method { (based on the cumulative
distribution function of the pixels in the TSZ map)} to determine
these thresholds, more sophisticated methods optimising the series of
{ TSZ} thresholds need to be investigated.

Using our straightforward choice of thresholds, we have investigated
two methods to reconstruct the final KSZ maps: { A decorrelation
and a minimisation.}  The first method is based on the decorrelation
approach using the PCA. { It significantly underestimates the
standard deviations of the reconstructed KSZ maps as compared to the
original signal} by 50\% on average. More sophisticated decorrelation
methods can also be used.  Preliminary tests with the Independent
Component Analysis (ICA) \cite[]{cardoso93,hyv99} give promising
results in terms of the standard deviations. However, the results
obtained from the ICA need to be rescaled using external flux
constraints which are not always (or easily) available in
``real-life''. For example in our case, we would need to use the
fluxes of known clusters to calibrate the reconstructed KSZ maps on
the original signal. Despite this limitation, we will continue
investigating this method in the future.  The decorrelation method is
a blind method which advantage is that no a priori criteria are needed
to obtain the KSZ map.  However, the resulting maps are of low quality
in terms of standard deviation. The second reconstruction method we
use is based on a minimisation technique that takes into account the
statistical properties of the KSZ signal, namely: (i) KSZ dominates
over the primary anisotropies at small angular scales, and (ii) the
KSZ fluctuations follow a non-Gaussian distribution with a non-zero
excess kurtosis. In the present study, we use the excess kurtosis of
the diagonal wavelet coefficients to characterise the non-Gaussian
signatures of the KSZ effect. However, we could generalise the
minimisation criterion to include the third moment (skewness) { in
order to} account for, and thus separate,
between different processes contributing to the signal and having
different non-Gaussian characters. The minimisation method we propose
gives reconstructed KSZ maps that are in quite good agreement with the
original signal with an average correlation coefficient between
original and reconstructed KSZ map of 0.78, and an error of 5\% in the
standard deviation of the reconstructed KSZ maps. However, the
minimisation method depend greatly on the minimisation criteria and
therefore on an a priori knowledge of the reconstructed signal.

The results presented here are based on an ideal case where only the
two signals CMB and SZ are taken into account. This simplified test
case allows us to investigate the ultimate intrinsic limitations of
the method.  The { investigation} of noise and additional
astrophysical contributions is quite important { but it is beyond
the scope of our present study}.  It should a priori be partly treated
in the first step component separation { (from which we obtain the
observables: $y$ and $\delta_T$ maps)}.  However, some contribution
from the TSZ signal may remain in the $\delta_T$ map, because of
imperfect component separation or when the relativistic corrections to
the SZ effect are not corrected for, for example. { This will act
as an additional and correlated noise.} As shown by \cite{diego03},
this introduces a non-Gaussian signature into the CMB signal and hence
errors in the KSZ reconstruction. This non-Gaussian contribution due
to the TSZ effect is characterised by a non-zero skewness. { We can
account for this source of correlated noise and thus correct for it},
either at the interpolation stage with the additional constraint that
the skewness should be zero { (which is the case for the primary
CMB anisotropies), or} in the minimisation procedure using a
generalised criterion including the skewness as well as the excess
kurtosis. Another way to overcome this difficulty, is to apply our
method to the individual frequency maps and correct for { any} TSZ
spurious contribution. As a matter of fact, the technique we present
can be applied to separate TSZ fluctuations from the primary
fluctuations.  The correlation between two frequency channels, where
TSZ dominates, gives indeed a first order spatial template which can
be used to obtain the TSZ signal and thus to predict the primary CMB.
At this point, the first order $y$ map can be used in the next step to
better estimate, in an iterative way, the TSZ signal itself. We plan
to investigate this method in the future.  As for the {
instrumental} noise, it can be taken into account in the interpolation
step by relaxing the parameter $\lambda$ (Eq. 2). However, if the
noise is not white then other interpolation methods might have to be
used { (Sec \ref{sec:inter}) for discussion}.  Another way to deal
with the noise is to minimise not on the non-Gaussian character of the
KSZ, but rather on the statistical properties of the remainder
(i.e. CMB+noise+other components) at scales where CMB dominates.  We
will then obtain an estimate of all the components except KSZ that can
then be subtracted to the total signal.

{ The present work is based on the use of a spatial template to separate
KSZ temperature fluctuations from the primary fluctuations.}  This
point has been already noted by \cite{haehnelt96} who used an X-ray
emission template to measure the peculiar velocity of clusters. {
The choice of the spatial template is an
important issue for our method since it is used to define the mask and
hence the interpolated regions. The spatial template should then be
the closest possible to the signal (SZ effect in our case). The
optimal choice is really to use the TSZ template itself (similarly to the
commonly used matched filter approach). Since SZ traces the intra-cluster gas,
we could also use the X-ray emission of clusters as a template.} The
problem in this case is that the  X-ray emission scales with the product
$n_{e}^{2}T_{e}^{1/2}$, whereas the TSZ scales with $n_{e}T_{e}$, and
consequently the spatial extension of clusters is underestimated by
taking X-ray templates. Additionally, { the X-ray observations of
galaxy clusters are restricted to a rather small fraction of objects
not too distant to suffer from the dimming effect and with high enough
intra-cluster temperatures to be detected. The TSZ effect on the
contrary is redshift independent and less sensitive to the gas
parameters. Moreover, using the TSZ map as a template in our method
has the advantage of evaluating the temperature fluctuations (even
very low amplitude ones) associated with KSZ in the map without
resorting to the knowledge or the measurement of the cluster
parameters ($n_{e}, T_{e}$).  The method presented here has proven its
success in achieving this goal.  In particular, Sec. \ref{sec:sens}
illustrates how well is the KSZ map reconstructed when the input KSZ
signal is decreased by { one} order of magnitude in terms of standard
deviation.} 

\section{Conclusion}

{ In this first attempt to extract a map of the KSZ temperature
fluctuations from the CMB anisotropies we use a method which is based
on very simple and minimal assumptions. We discuss the issue of noise
and astrophysical contributions but we do not take them explicitly
into account. Therefore, our results show the intrinsic
limitations of the method in terms of reconstructing a KSZ map from a
mixture of CMB and KSZ anisotropies. We demonstrate that the 15 KSZ
reconstructed maps are in quite good agreement with the original input
signal with a correlation coefficient between original and
reconstructed maps of 0.78 on average, and an error on the standard
deviation of the reconstructed KSZ map of only 5\% on average.

To achieve these results, we use the hypothesis that a first step
component separation provides us with: (i) a map of Compton parameters
for the TSZ effect of galaxy clusters, and (ii) a map of temperature
fluctuations for the primary CMB + KSZ cluster signal. Our method
essentially takes benefit from the spatial correlation between KSZ and
TSZ effects towards the same galaxy clusters. This correlation allows
us to use the TSZ map as a spatial template in order to mask, in the
CMB + KSZ map, the pixels where the clusters must have imprinted an SZ
fluctuation. In practice a series of TSZ thresholds is defined and for
each threshold, we estimate the corresponding KSZ signal by
interpolating the CMB fluctuations on the masked pixels.  The series
of estimated KSZ maps finally is used to reconstruct the KSZ map
through the minimisation of a criterion taking into account two
statistical properties of the KSZ signal (KSZ dominates over the
primary anisotropies at small scales, KSZ fluctuations are
non-Gaussian distributed).} 

\begin{acknowledgements}
The authors would like to thank Simon Masnou, { Jose-Luis Sanz} and
Philippe Th\'evenaz for fruitful discussions. We also wish to warmly
thank Steen Hansen and Jean-Loup Puget for their helpful comments {
and an anonymous referee for his/her remarks on an earlier version}.
\end{acknowledgements}

\bibliographystyle{natbib} 

\begin{thebibliography}{}

\bibitem[{Aghanim} et~al.(1997)]{aghanim97}
{Aghanim}, N., {De Luca}, A., {Bouchet}, F.~R., {Gispert}, R. \& {Puget}, J.~L.
\newblock 1997, {\em \AaA}, 325, 9.
\bibitem[{Aghanim} \& {Forni}(1999)]{aghanim99} {Aghanim}, N. \& {Forni}, O.
\newblock 1999, {\em \AaA}, 347, 409.
\bibitem[{Aghanim} et~al.(2001)]{aghanim01}
{Aghanim}, N., {Gorski}, K.~M., \& {Pujet}, J.-L.
\newblock 2001, {\em \AaA}, 374, 1
\bibitem[{Baccigalupi} et~al.(2000)]{baccigalupi00} {Baccigalupi} C. et~al
\newblock 2000, {\em \MNRAS}, 318, 769.
{ \bibitem[{Benson} et~al.(2003)]{benson03} {Benson}, B.~A., {Church}, S.~E., 
{Ade}, P.~A. et~al. \newblock 2003  {\em ApJ}, 592, 674.}
\bibitem[{Birkinshaw}(1999)]{birkinshaw99} {Birkinshaw} M.
\newblock 1999, {\em Phys. Reports}, 310, 97.
\bibitem[{Bouchet} \& {Gispert}(1999)]{bouchet99}
{Bouchet}, F.~R. \& {Gispert}, R. \newblock 1999, {\em New Astron.}, 4, 443.
\bibitem[{Cardoso} \& {Soulamiac}(1993)]{cardoso93}
{Cardoso}, J.~-F. \& {Soulamiac}, A.
\newblock 1993, {\em IEE Proceedings-F}, 140, 362.
{ \bibitem[{Carlstrom} et~al.(2002)]{carlstrom02}
{Carlstrom}, J.E., {Holder}, G.P. \& {Reese}, E.D.
\newblock 2002, {\em ARA\&A}, 40, 643.}
\bibitem[{Cayon} et~al.(2003)]{cayon03}
{Cayon}, L., {Martinez-Gonzalez}, E., {Argueso}, F., {Banday}, A.~J. \&
  {G\'orski}, K.~M. \newblock 2003, {\em \MNRAS}, 339, 1189
\bibitem[{Cohen} et~al.(1990)]{cohen90}
{Cohen}, A., {Daubechies}, I. \& {Feauveau}, J.~C.
\newblock 1990, Technical report, AT\&T Bell Lab., Page:TM 11217-900529-07.
{ \bibitem[{Delabrouille} et~al.(2002)]{delab02}
Delabrouille J., Cardoso J.-F. \& Patanchon G.
\newblock 2002, {\em MNRAS, submitted}, astro-ph/0211504}
{ \bibitem[{Diaferio} et~al.(2000)]{diaferio00}
Diaferio, A.,  Sunyaev R.~A. \& Nusser A. 
\newblock 2000, {\em ApJ}, 533, 7.}
\bibitem[{Diego} et~al.(2003)]{diego03}
{Diego}, J.~M., Hansen, S.~H. \& Silk, J.
\newblock 2003, {\em MNRAS}, 338, 796 \bibitem[Faber \& Tully(1976)]{faber76}
{Faber}, S.~M \& {Jackson}, R.~E. \newblock 1976, {\em \ApJ}, 204, 668.
\bibitem[{Forni} \& {Aghanim}(1999)]{forni99} {Forni}, O. \& {Aghanim}, N.
\newblock 1999, {\em \AaAS}, 137, 553.
\bibitem[{Haehnelt} \& {Tegmark}(1996)]{haehnelt96}
{Haehnelt}, M.~G \& {Tegmark}, M. \newblock 1996, {\em \MNRAS}, 279, 545.
\bibitem[{Hobson} et~al.(1998)]{hobson98}
{Hobson} M.~P., {Jones} A.~W., {Lasenby} A.~N. \& {Bouchet} F.~R.
\newblock 1998, {\em \MNRAS}, 300, 1.
{ \bibitem[{Hobson \& McLachlan}(2003)]{hobson03} 
{Hobson} M.~P. \& McLachlan C.
\newblock 2003, {\em MNRAS}, 338, 765.}
\bibitem[{Holzapfel} et~al.(1997)]{holzapfel97} {Holzapfel} W.~L., et~al.
\newblock 1997, {\em ApJ}, 481, 35 
\bibitem[{Hyv\"arinen}(1999)]{hyv99} {Hyv\"arinen} A.
\newblock 1999 {\em IEEE Transactions on Neural Networks}, 10, 626.
\bibitem[{Komatsu} et~al.(2003)]{komatsu03}
{Komatsu}, E., {Kogut}, A., {Nolta}, M.~R. et~al.
\newblock 2003, {\em ApJ}, 148, 119
\bibitem[{Kuo} et~al.(2002)]{kuo02} Kuo, C.~L., et~al.
\newblock 2002, {\em ApJ}, submitted, astro-ph/0212289
\bibitem[{Lamarre} et~al.(1998)]{lamarre98} {Lamarre}, J.~-M et~al.
\newblock 1998, {\em ApJ}, 566, 19.
\bibitem[{Maisinger} et~al.(2003)]{maisinger03}
Maisinger, K., {Hobson} M.~P.  \& {Lasenby} A.~N.
\newblock 2003, {\em \MNRAS}, submitted, astro-ph/0303246
\bibitem[{Ostriker} \& {Vishniac}(1986)]{ostriker86}
{Ostriker}, J.~P. \& {Vishniac}, E.~T. \newblock 1986, {\em \ApJ}, 306, L51.
\bibitem[{Press} et~al.(1992)]{press92}
Press, W.~H., Teukolsky, S.~A., Vetterling, W.~T.  \& Flannery, B.~P.
\newblock 1992, {\em Numerical Recipes in FORTRAN: The Art of Scientific Computing},
Cambridge University Press. 
\bibitem[{Rees} \& {Sciama}(1968)]{rees68}
{Rees}, M.~J. \& {Sciama}, D.~W. \newblock 1968, {\em \Natur}, 511, 611.
{ \bibitem[{Reeves}(1994)]{reeves94} {Reeves} S.~J.
\newblock 1994, {\em IEEE Transactions Image Processing}, 3, 319.}
\bibitem[{Rephaeli}(1995)]{rephaeli95} {Rephaeli}, Y.
\newblock 1995, {\em Ann. Rev. Astron. \& Astrophys.}, 445, 33.
\bibitem[{Santos} et~al.(2003)]{santos03} {Santos}, M.~G., et al.
\newblock 2003, {\em \MNRAS}, 341, 623.
{ \bibitem[{Sorel} et~al.(2002)]{sorel02} {Sorel}, M., {Aghanim} N. \& {Forni} O.
\newblock 2002, {\em A\&A}, 395, 747.}
\bibitem[{da Silva} et~al.(2001)]{dasilva01}
{da Silva}, A.~C., {Barbosa}, D., {Liddle}, A.~R. \& {Thomas}, P.~A.
\newblock 2001, {\em \MNRAS}, 326, 155.
\bibitem[{Sunyaev} \& {Zel'dovich}(1972)]{suniaev72}
{Sunyaev}, R.~A. \& {Zel'dovich}, I.~B.
\newblock 1972, {\em Comments Astrophys. Space Phys.}, 4, 173.
\bibitem[{Sunyaev} \& {Zel'dovich}(1980)]{suniaev80}
{Sunyaev}, R.~A. \& {Zel'dovich}, I.~B. \newblock 1980, {\em \ARAA}, 18, 537.
\bibitem[{Starck} et~al.(2003)]{starck03}
{Starck}, J.~L., {Aghanim}, N. \& {Forni}, O.
\newblock 2003, {\em \AaA}, accepted for publication. 
\bibitem[Th\'evenaz et~al.(2001)]{thevenaz01}
{Th\'evenaz}, P, {Blu}, T. \& {Unser}, M. 
\newblock 2001, {\em IEEE Transactions on Medical Imaging}, 19, 739.
\bibitem[Tully \& Fisher(1977)]{tully77} {Tully}, R.~B \& {Fisher}, J.~R.
\newblock 1977, {\em \AaA}, 54, 661. 
\bibitem[Unser et~al.(1993)]{unser93}
{Unser}, M., {Aldroubi}, A. \& {Eden} M.
\newblock 1993, {\em IEEE Transactions on Signal Processing}, 41, 834.
\bibitem[Unser(1995)]{unser95} {Unser}, M. \newblock 1995,
{\em Proceedings of the 1995 IEEE International Conference on Image Processing}, 1, 49.
{ \bibitem[{Vielva} et~al.(2003)]{vielva03}
{Vielva} P., {Martinez-Gonzalez} E., {Barreiro} R.~B. et~al
\newblock 2003,  astro-ph/0310273   }
\bibitem[{Villasenor} et~al.(1995)]{villasenor95}
{Villasenor}, J.~D., {Belzer}, B. \& {Liao}, J.
\newblock 1995, {\em IEEE Transactions on Image Processing}, 4, 1053.
\bibitem[{Vishniac}(1987)]{vishniac87} {Vishniac}, E.~T.
\newblock 1987, {\em \ApJ}, 322, 597. 
{ \bibitem[{Wahba}(1977)]{wahba77} {Wahba}, G.
\newblock 1977, {\em SIAM J. Math. Anal.} 14, 651.} 
\bibitem[{Wesseling}(1992)]{wesseling92} {Wesseling}, P.
\newblock 1992, {\em An Introduction to Multigrid Methods}, J. Wiley \& Sons.

\end{thebibliography}
 
\end{document}